\documentclass[%
twocolumn,
10pt,
superscriptaddress,
amsmath,amssymb,
aps,
pra,
floatfix,
]{revtex4-2}

\usepackage{graphicx}
\usepackage{bm}
\usepackage{times}
\usepackage{amsmath,amssymb}
\usepackage{amsthm}
\usepackage{enumerate}
\usepackage{multirow}
\usepackage{siunitx}
\usepackage{makecell}
\usepackage[caption=false]{subfig}
\usepackage{tikz}
\usetikzlibrary{quantikz2}
\usepackage{tabularx}

\usepackage[colorlinks=true,linkcolor=blue,citecolor=red, linktocpage=true,breaklinks=true]{hyperref}

\newcommand{\eq}{\begin{equation}}
\newcommand{\en}{\end{equation}}
\newcommand{\eqa}{\begin{eqnarray}}
\newcommand{\ena}{\end{eqnarray}}

\newcommand{\pr}{\mathrm{Pr}}

\begin{document}

\title{Near-deterministic quantum search algorithm without phase design}

\author{Zhen \surname{Wang}}
\affiliation{School of Mechanical and Material Engineering, Xi'an University, Xi’an 710065, China}

\author{Kun \surname{Zhang}}
\email{kunzhang@nwu.edu.cn}
\affiliation{School of Physics, Northwest University, Xi’an 710127, China}
\affiliation{Shaanxi Key Laboratory for Theoretical Physics Frontiers, Xi'an 710127, China}
\affiliation{Peng Huanwu Center for Fundamental Theory, Xi'an 710127, China}

\author{Vladimir \surname{Korepin}}
\affiliation{C.N. Yang Institute for Theoretical Physics, Stony Brook University, New York 11794, USA}

\date{\today}

\begin{abstract}

Grover's algorithm solves the unstructured search problem. Grover's algorithm can find the target state with certainty only if searching one out of four. Designing the deterministic search algorithm can avoid any repetition of the algorithm, especially when Grover's algorithm is a subroutine in other algorithms. Grover's algorithm can be deterministic if the phase of the oracle or the diffusion operator is delicately designed. The precision of the phases could be a problem. A near-deterministic quantum search algorithm without the phase design is proposed. The algorithm has the same oracle and diffusion operators as Grover's algorithm. One additional component is the rescaled diffusion operator. It acts partially on the database. The success probability of Grover's algorithm is improved by the partial diffusion operator in two different ways. The possible cost is one or two more queries to the oracle. The deterministic search algorithm is also designed when searching one out of eight, sixteen, and thirty-two.

\end{abstract}

\maketitle

\section{\label{sec:intro} Introduction}

The unstructured search problem can be solved by exhaustively examining each item in the database. The complexity is $\mathcal O(N)$ with $N$ as the number of items in the database. Grover's algorithm is one of the major achievements in quantum computation. It solves the unstructured search problem by ``examining'' $\mathcal O(\sqrt{N})$ times of the database \cite{grover1996fast,grover1997quantum}. Examining means evaluating the one-way function, also known as the oracle, which can identify the target state. Grover's algorithm is optimal in the number of queries to the oracle \cite{BBHT98,Zalka99}. 

Quantum mechanics is probabilistic. Therefore most quantum algorithms are probabilistic, also true for Grover's algorithm. Grover's algorithm starts with the equal superposition of $N=2^n$ computational basis states ($n$ qubits). Each basis state represents a search item. The amplitude of the target state is amplified by the Grover operator. The Grover operator is composed of the oracle operator (query to the oracle) and the diffusion operator (mixing the amplitudes of target and non-target states). Eventually, the amplitude of the target state approaches to 1. Therefore the measurement can reveal the solution with high probabilities. Algorithms in such a framework are called amplitude amplification \cite{grover1998quantum,brassard2002quantum}. 

Grover's algorithm has a simple geometric picture. It rotates the initial state to the target state on a two-dimensional plane, namely the $SO(2)$ picture. Suppose that the number of target states is $M$. Define the ratio $\lambda = M/N$. The optimal number of the Grover operator, which gives the maximal success probability, is given by $k_\text{opt} = \lfloor \pi/(4\theta_\lambda)-1/2\rceil$ with $\theta_\lambda = \arccos(1/\sqrt{\lambda})$. Because of the nature of trigonometric functions, Grover's algorithm is certain only if the ratio between the number of target states and the total number of states $N$ is $1/4$ \cite{diao2010exactness}. When the ratio $\lambda$ is relatively large, such as $\lambda>10^{-1}$, the probability of Grover's search becomes an issue. Several modified Grover's algorithms are proposed to have the deterministic search \cite{brassard2002quantum,hoyer2000arbitrary,long2001grover,roy2022deterministic,leng2023improving}. They are all based on designing phases in the oracle operator or the diffusion operator. Then the $SO(2)$ picture is extended to the $SO(3)$. The rotation angle and axis are determined by the phase. Therefore one can always design the perfect match between the final state and the target state. Besides, the phase design is also applied to other variants of Grover's algorithm, such as the fixed-point search algorithm \cite{yoder2014fixed}, the variational-like search algorithm \cite{morales2018variational}, and the QAOA-like search algorithm \cite{jiang2017near}.

The delicately designed phase in the deterministic search algorithm requires very high precision. It could be a challenge for practice. There is another way to invoke the three-dimensional picture of the quantum search algorithm. It is realized by introducing the rescaled diffusion operator. We call it the partial or local diffusion operator in the following. Correspondingly, the original diffusion operator is called the global diffusion operator. Quantum search algorithm with both the global and local diffusion operators has the $O(3)$ picture \cite{korepin2006group}. Grover operator with the partial diffusion operator realizes a renormalized search in the subspace of the database. It was introduced in designing the quantum partial search algorithm, which trades accuracy for speed \cite{GR05,KG06,Korepin05}. Also because of the smaller circuit depth of the local diffusion operator, it was introduced to optimize the depth of the quantum search algorithm \cite{grover2002trade,zhang2020depth,brianski2021introducing}. 

In this work, we study how to improve the success probability of the search algorithm via the local diffusion operator. Because of the nature of the local diffusion operator, we show that the algorithm can have mid-circuit measurements. In other words, the search algorithm can be designed into two steps or two stages. As examples, we design the deterministic search algorithm with $n=3,4,5$ and the near-deterministic search algorithm with $n=6,7,8,9$ (assuming that there is only one target state). Here the near-deterministic means that the success probability is above $99.9\%$. When $n$ is relatively small, the success probability of Grover's algorithm is not near-deterministic (below $99.9\%$). As $n$ increases, Grover's algorithm becomes near-deterministic. If there are multiple target states, our strategy also works, however requires additional information on the distribution of the target states, similar to the quantum partial search algorithm \cite{GR05,KG06,Korepin05}.

The paper is organized as follows. Sec. \ref{sec:quantum_search} reviews Grover's algorithm and quantum partial search algorithm. Secs. \ref{sec:one_stage} and \ref{sec:two_stage} show the improved one-stage and two-stage quantum search algorithms respectively. We also provide the $O(3)$ picture of the algorithm. We discuss the multi-target cases in Sec. \ref{sec:multi-target}. The final section is the conclusion. 

\section{\label{sec:quantum_search} Quantum search algorithm}

We first review Grover's algorithm in Subsection \ref{subsec:Grover}. Our near-deterministic search algorithm is based on the techniques from the quantum partial search algorithm, of which we give a brief explanation in Subsection \ref{sub:QPS}.

\subsection{\label{subsec:Grover}Grover's algorithm}

The advantage of quantum computation is parallelism due to the superposition. The initial state of Grover's algorithm is the equal superposition of $N=2^n$ basis states, denoted as
\begin{equation}
    |s_n\rangle = \frac{1}{\sqrt{2^n}}\sum_{j=0}^{2^n-1}|j\rangle.
\end{equation}
It is a maximal coherent state but can be simply constructed from the single-qubit Hadamard gate \cite{nielsenQuantumComputationQuantum2010}. The oracle can recognize the target state $t$, given by $f(t)=1$ and $f(x\neq t)=0$. We assume that there is a unique target state in the database. We discuss the multi-target cases in Sec. \ref{sec:multi-target}. Applying the trick of phase kickback, the equivalent form of the oracle operator is
\begin{equation}
\label{eq:O_t}
    O_t = 1\!\!1_{2^n}-2|t\rangle\langle t|,
\end{equation}
with the identity operator of $\mathcal H_2^{\otimes n}$ ($2^n \times 2^n$ identity matrix). It reflects the state in terms of the target state $|t\rangle$. The diffusion operator is another reflection operator, given by
\begin{equation}
\label{eq:D_n}
    D_n = 1\!\!1_{2^n}-2|s_n\rangle\langle s_n|.
\end{equation}
It reflects in terms of $|s_n\rangle$. Since $|s_n\rangle$ is the equal superposition of the basis states, the reflection is around the average amplitude of the states. It can be realized from the multi-qubit Toffoli gate \cite{nielsenQuantumComputationQuantum2010}. The phase design of $O_t$ or $D_n$ is to have a phase in the reflection instead of the fixed value $-1$.

Grover operator is the product of $O_t$ and $D_n$, namely
\begin{equation}
\label{eq:G_n}
    G_n = -D_nO_t.
\end{equation}
The minus sign does not influence the success probabilities. It is only convenient when considering the geometric picture. The action of $G_n$ on the initial state $|s_n\rangle$ does not distinguish the amplitude of non-target states. Therefore we can group them as
\begin{equation}
\label{eq:t_perp}
    |t^\perp\rangle = \frac{1}{\sqrt{2^n-1}}\sum_{j\neq t}|j\rangle.
\end{equation}
Naturally, we have $\langle t|t^\perp\rangle = 0$. The initial state $|s_n\rangle$ can be rewritten as
\begin{equation}
    |s_n\rangle = \sin\theta_n|t\rangle+\cos\theta_n|t^\perp\rangle.
\end{equation}
The angle $\theta_{n}$ is given by $\theta_n = \arcsin(1/\sqrt{2^n})$. In the orthonormal basis $\{|t\rangle,|t^\perp\rangle\}$, the Grover operator $G_n$ is a $2\times 2$ orthogonal matrix with determinant 1. Therefore the geometric picture is natural due to $G_n\in SO(2)$. Grover operator rotates the initial state to the target state in the manner
\begin{equation}
\label{eq:G_n^k}
    G_n^k|s_n\rangle = \sin\left((2k+1)\theta_n\right)|t\rangle + \cos\left((2k+1)\theta_n\right)|t^\perp\rangle.
\end{equation}
The success probability finding the target state $|t\rangle$ approaches to 1 if $(2k+1)\theta_n$ approaches to $\pi/2$. The optimal iteration number $k$ is
\begin{equation}
    k_\text{opt} = \left\lfloor\frac{\pi}{4\theta_n} - \frac  1 2\right\rceil.
\end{equation}
Here $\lfloor\cdot\rceil$ denotes the closest integer. Since $k_\text{opt}$ is an integer, the term $(2k_\text{opt}+1)\theta_n$ can be exact to $\pi/2$ only if $n=2$ \cite{diao2010exactness}. In other cases, Grover's algorithm is not deterministic. For example, consider $n=4$ which gives $k_\text{opt}=3$. The maximal success probability is around $96.13\%$.

\subsection{\label{sub:QPS}Quantum partial search algorithm}

The $n$-qubit search means that the bit length of the solution $t$ is $n$, represented by the $n$-qubit basis state $|t\rangle$. We can divide the target state into two parts, namely $|t\rangle = |t_1\rangle\otimes|t_2\rangle$. Suppose that the length of $t_2$ is $m$, then $t_1$ has the bit length $n-m$. Quantum partial search algorithm finds $t_1$ instead of the full solution $t$ \cite{GR05,KG06,Korepin05}. It is realized with the help of the rescaled diffusion operator defined as
\begin{equation}
\label{eq:D_nm}
    D_{n,m} = 1\!\!1_{2^{n-m}}\otimes \left(1\!\!1_{2^m}-2|s_m\rangle\langle s_m|\right).
\end{equation}
We also call $D_{n,m}$ as the local or partial diffusion operator. It implements the reflection on the average on $m\leq n$ qubits. Correspondingly, we have the Grover operator
\begin{equation}
    G_{n,m} = -D_{n,m}O_t.
\end{equation}
It is called the local or partial Grover operator. For simplicity, we denote $G_m=G_{n,m}$ when there is no ambiguity. The local Grover operator $G_m$ has the quantum circuit diagram
\begin{equation}
\begin{quantikz}[align equals at=1.5]
& \gate[2]{G_m} & \qwbundle{n-m}\\
&  & \qwbundle{m} \\
\end{quantikz}=\hspace{2mm}\begin{quantikz}[align equals at=1.5]
 & \gate[2]{O_t} &  & \qwbundle{n-m} \\
 &  & \gate{D_m} & \qwbundle{m} 
\end{quantikz}
\vspace{-.5cm}
\end{equation}
We use the \verb!quantikz! package drawing the quantum circuit diagram \cite{kay2018tutorial}. 

The local Grover operator can amplify the amplitude of the target state. But it leaves some non-target states unchanged. Specifically, consider the normalized state \footnote{We take the notation from the quantum partial search algorithm. Here ``ntt'' stands for the non-target states in the target block.}
\begin{equation}
    |ntt\rangle = |t_1\rangle\otimes\frac{1}{\sqrt{2^m-1}}\sum_{j\neq t_2}|j\rangle.
\end{equation}
Then the local Grover operator works as
\begin{multline}
    G_m^k|t_1\rangle\otimes|s_m\rangle \\
    = \sin\left((2k+1)\theta_m\right)|t\rangle + \cos\left((2k+1)\theta_m\right)|ntt\rangle.
\end{multline}
It is a Grover search on the subspace spanned by $m$ qubits. However, $G_m$ does not change the amplitudes of all non-target states as $G_n$ does. Exclude the non-target state $|ntt\rangle$, then the rest (normalized) non-target state is
\begin{equation}
    |u\rangle = \frac{1}{\sqrt{2^{n-m}}}\left(\sqrt{2^n}|s_n\rangle - |t\rangle - \sqrt{2^m-1}|ntt\rangle\right).
\end{equation}
It is the normalized state that the beginning $n-m$ qubits are not $|t_1\rangle$. 

In the orthonormal basis $\{|t\rangle,|ntt\rangle,|u\rangle\}$, the initial state $|s_n\rangle$ can be rewritten as
\begin{multline}
    |s_n\rangle = \sin\theta_{n-m}\sin\theta_m|t\rangle+\sin\theta_{n-m}\cos\theta_m|ntt\rangle \\
    + \cos\theta_{n-m} |u\rangle.
\end{multline}
Note that $\theta_m = \arcsin(1/\sqrt{2^{m}})$ and $\theta_{n-m} = \arcsin(1/\sqrt{2^{n-m}})$. Correspondingly, we can write $G_n$ as a $3\times 3$ orthogonal matrix, given by
\begin{widetext}
\begin{equation}
    G_n = \begin{pmatrix}
        1- 2\sin^2\theta_{n-m}\sin^2\theta_m & 2\sin^2\theta_{n-m}\sin\theta_m\cos\theta_m & 2\sin\theta_{n-m}\cos\theta_{n-m}\sin\theta_m \\
        -2\sin^2\theta_{n-m}\sin\theta_m\cos\theta_m & 2\sin^2\theta_{n-m}\cos^2\theta_m - 1 & 2\sin\theta_{n-m}\cos\theta_{n-m}\cos\theta_m \\
        -2\sin\theta_{n-m}\cos\theta_{n-m}\sin\theta_m & 2\sin\theta_{n-m}\cos\theta_{n-m}\cos\theta_m & 2\cos^2\theta_{n-m} - 1 \\
        \end{pmatrix}.
\end{equation}
\end{widetext}
The determinant of $G_n$ is $-1$. Therefore it is an element of $O(3)$, instead of $SO(3)$ \cite{korepin2006group}. The global Grover operator $G_n$ is the product of two reflections in terms of $|t\rangle$ and $|s_n\rangle$. However, if the state is restricted in the plane spanned by $|t\rangle$ and $|s_n\rangle$, it is a valid rotation (as in Grover's algorithm). It implies that $G_n$ has an eigenvector with eigenvalue $-1$, which is perpendicular to the plane spanned by $|t\rangle$ and $|s_n\rangle$. 

\begin{figure*}
\subfloat{%
  \includegraphics[width=0.33\textwidth]{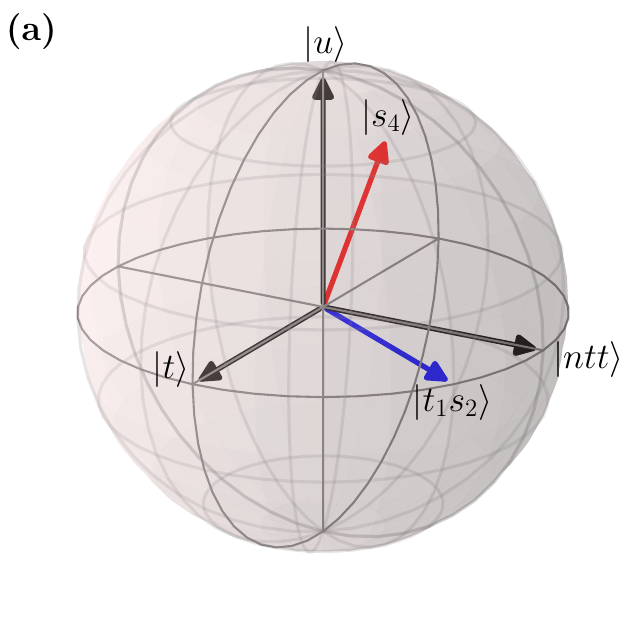}%
}\hfill
\subfloat{%
  \includegraphics[width=0.33\textwidth]{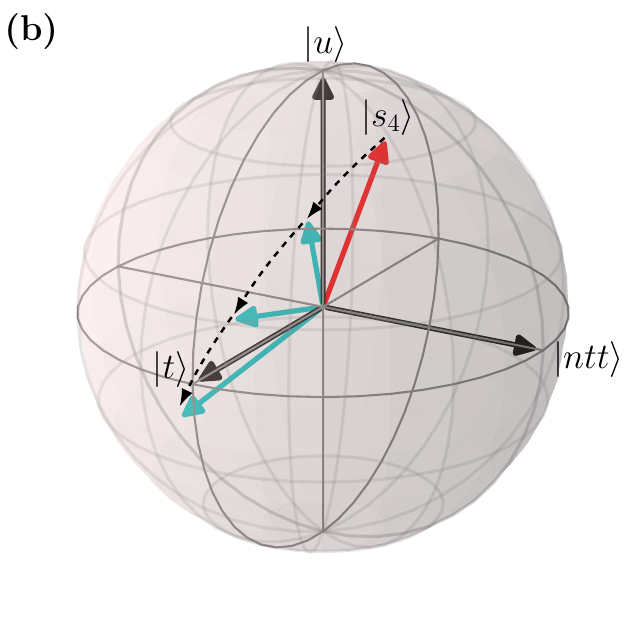}%
}\hfill
\subfloat{%
  \includegraphics[width=0.33\textwidth]{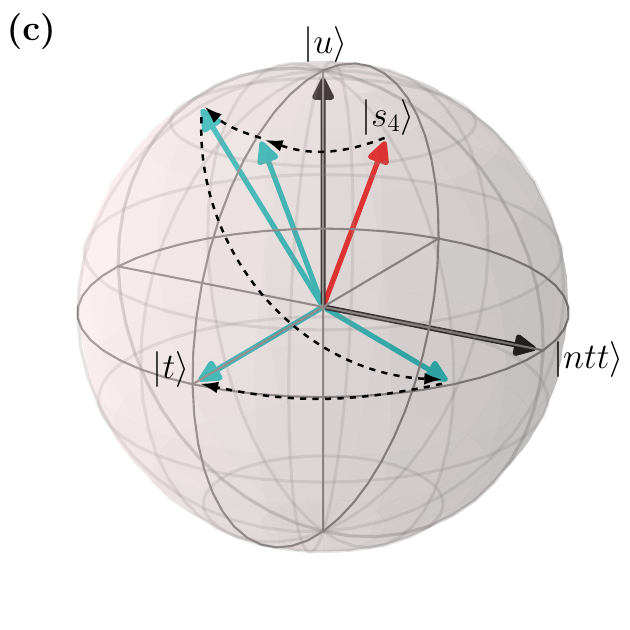}%
}
\caption{Geometric picture of $n=4$ quantum search algorithm with local diffusion operator. (a) The oracle operator and the global diffusion operator reflect the state in terms of $|t\rangle$ and $|s_4\rangle$ respectively. The local diffusion operator $D_2$ reflects in terms of $|s_2\rangle$, where $|t_1s_2\rangle = |t_1\rangle\otimes |s_2\rangle$ is on the plane spanned by $|t\rangle$ and $|ntt\rangle$. (b) Grover's algorithm rotates the initial state $|s_4\rangle$ to the target state $|t\rangle$. The final state is not perfectly matched with the target state. (c) The search algorithm is realized by $G_2G_4G_2^2$. The local Grover operator rotates the state around the axis $|u\rangle$. The global Grover operator reflects the state onto the plane spanned by $|t\rangle$ and $
|ntt\rangle$. The final state is perfectly matched with the target state. }\label{fig:4q}
\end{figure*}

The local Grover operator $G_m$ acts trivially on $|u\rangle$, namely $G_m|u\rangle = |u\rangle$. Therefore it is block diagonal in the basis $\{|t\rangle,|ntt\rangle,|u\rangle\}$. Specifically, we have
\begin{equation}
    G_m = \begin{pmatrix}
        \cos2\theta_m & \sin2\theta_m & 0 \\
        -\sin2\theta_m & \cos2\theta_m & 0 \\
        0 & 0 & 1 \\
    \end{pmatrix}.
\end{equation}
It is an element of the $SO(3)$ group. The rotation angle is $2\theta_m$. The rotation axis is $|u\rangle$. See Fig. \ref{fig:4q} for the example on $n=4$. Note that $G_n$ and $G_m$ do not commute (with $m\neq n$), therefore different orders acting on the initial state would give different success probabilities.

In the context of the quantum partial search algorithm, we can view that the database is divided into $2^{n-m}$ blocks. Each block has $2^m$ items. The partial solution $t_1$ is the block index of the target state. The algorithm is realized via the operator $G_nG_m^{k_2}G_n^{k_1}$ \cite{GR05}. Here the iteration number $k_1+k_2$ is minimized, at the condition that the amplitude of $|u\rangle$ is zero \cite{KG06,Korepin05}. The algorithm has the oracular complexity $\mathcal O(\sqrt{N}-\alpha\sqrt{b})$ with $b=2^m$ and positive coefficient $\alpha$. It is faster than Grover's algorithm, at the cost of accuracy.

\section{\label{sec:one_stage} One-stage near-deterministic quantum search algorithm}

The naive application of the partial diffusion operator is to search for a recalled database. We discuss such a strategy in Subsection \ref{subsec:nleq5}. In Subsection \ref{subsec:ngeq5}, we present a universal strategy to design the near-deterministic search algorithm via the partial diffusion operator. 

\subsection{\label{subsec:nleq5} Deterministic quantum search algorithm with \texorpdfstring{$n\leq 5$}{Lg}}

Grover algorithm is exact if $n=2$. It only requires one $G_2$ operator. Grover's algorithm with $n=3$ finds the target state with the success probability $94.53\%$ by $G^2_3$. We can take advantage of the exactness of the $n=2$ case for designing the $n=3$ algorithm. Specifically, one operator $G_{3,2}$ can find the target state with $100\%$ probability if the first qubit (without acting by the diffusion operator) is the $t_1$. Therefore, we can randomly guess $t_1\in\{0,1\}$. The circuit
$$
\begin{quantikz}
 \lstick{\ket{0}} & \gate{X^{a_1}} & \gate[3]{O_t} & & \\
 \lstick{\ket{0}} & \gate{H} & & \gate[2]{D_2} & \meter{} \\
 \lstick{\ket{0}} & \gate{H} & & & \meter{} 
\end{quantikz}
$$
finds the target state for certain if $a_1=t_1$. Suppose that the measurement result is $a_2a_3$. We can verify $a_1a_2a_3$ via one query to the classical oracle. If it is not the solution, then we do the above circuit again with $a = a\oplus 1$. It gives the target state for certain. In total, we can deterministically find the solution with two queries to the quantum oracle. The above strategy can also be implemented in parallel. It is also called the multi-programming of quantum search algorithm \cite{park2023quantum}. We omit the resource of classical oracle since it is much less valuable than quantum resources. 

The $4$-qubit Grover's algorithm has the optimal iteration number $k_\text{opt} = 3$. If we apply the above deterministic multi-programming strategy, the worst case requires four oracle operators (with four random guesses of two qubits). Interestingly, we find another one-stage deterministic search of $n=4$, given by
\begin{equation}
\label{eq:4q_deterministic}
    |\langle t|G_2G_4G_2^2|s_4\rangle|^2 = 1.
\end{equation}
The geometric picture based on the basis $\{|t\rangle,|ntt\rangle,|u\rangle\}$ can be found in Fig. \ref{fig:4q}. The middle global operator $G_4$ reflects the state onto the plane spanned by $|t\rangle$ and $|ntt\rangle$. 

We can design the deterministic 5-qubit search algorithm by random guessing three qubits (based on the one-stage deterministic 2-qubit search algorithm) or one qubit (based on the one-stage deterministic 4-qubit search algorithm). The total oracle number in both cases is eight. Although deterministic, it doubles the oracle number compared to Grover's algorithm with $k_\text{opt}=4$. We can design a more efficient near-deterministic (success probability above $99.9\%$) 5-qubit search algorithm with one or two more oracle operators. See Sec. \ref{sec:two_stage}. 

\subsection{\label{subsec:ngeq5} Near-deterministic quantum search algorithm with \texorpdfstring{$n \geq 6$}{Lg}}

\begin{figure}
    \centering
    \includegraphics[width=\columnwidth]{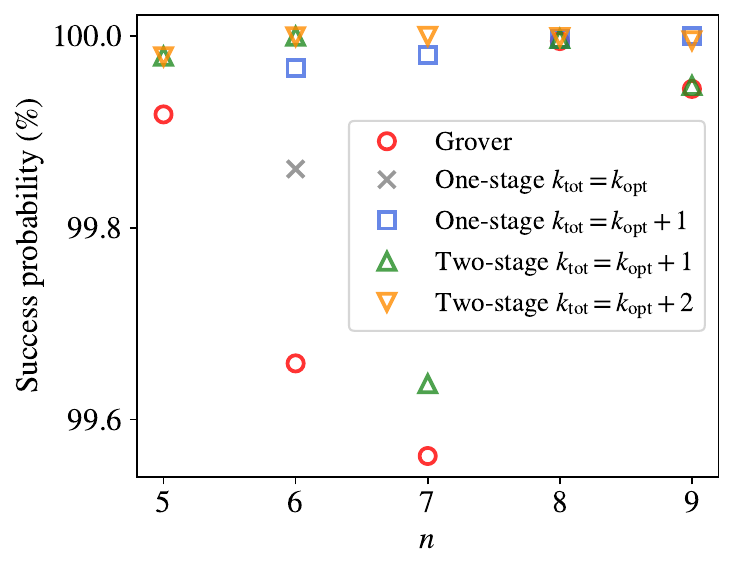}
    \caption{Success probabilities of designed one-stage and two-stage search algorithms. The total number of oracle operators is limited to one or two additional to Grover's algorithm with $k_\text{opt}$. The explicit expressions of the search operators are listed in Tables \ref{tab:one_stage} and \ref{tab:two_stage}.}
    \label{fig:success_probability}
\end{figure}

\begin{table*}[t]
    \caption{Comparison between Grover's algorithm and one-stage near-deterministic search algorithm. The designed operator is limited to $k_\text{tot} = k_\text{opt}$ or $k_\text{tot} = k_\text{opt}+1$. If no operator better than Grover's algorithm is found, it is listed as NA. The notation $\#$operator means the number of operators giving the success probability higher than Grover's algorithm. Specifically, the numerical criterion is that the success probability is larger than Grover's algorithm at least $10^{-4}\%$.}
    \centering
    \begin{tabular}{c|c|c|c|c|c|c|c|c}
    \hline\hline
         \multirow{2}{*}{$~~n~~$} & \multicolumn{2}{c|}{Grover} & \multicolumn{3}{c|}{One-stage $k_\text{tot}=k_\text{opt}$} & \multicolumn{3}{c}{One-stage $k_\text{tot}=k_\text{opt}+1$} \\\cline{2-9}
         & $~~k_\text{opt}~~$ & Pr (\%) & Pr (\%) & Operator & ~~\#Operator~~ & Pr (\%) & Operator & ~~\#Operator~~ \\\hline
         6 & 6 & ~~99.65857~~ & ~~99.86130~~ & $~~S_{6,5}(1,1,1,2,1)~~$ & 5 & ~~99.96643~~ & $S_{6,3}(1,1,2,1,2)$ & 1 \\
         7 & 8 & 99.56199 & NA & NA & NA & 99.98000 & $S_{7,5}(2,1,2,2,1,1,0)$ & 31 \\
         8 & 12 & 99.99470 & NA & NA & NA & 99.99724 & $~~S_{8,5}(2,1,3,1,2,1,2,1,0)~~$ & 5 \\
         9 & 17 & 99.94480 & NA & NA & NA & 99.99998 & $S_{9,6}(1,1,2,1,2,7,4)$ & 5047 \\
    \hline\hline
    \end{tabular}
    \label{tab:one_stage}
\end{table*}

\begin{figure*}
\subfloat{%
  \includegraphics[width=0.33\textwidth]{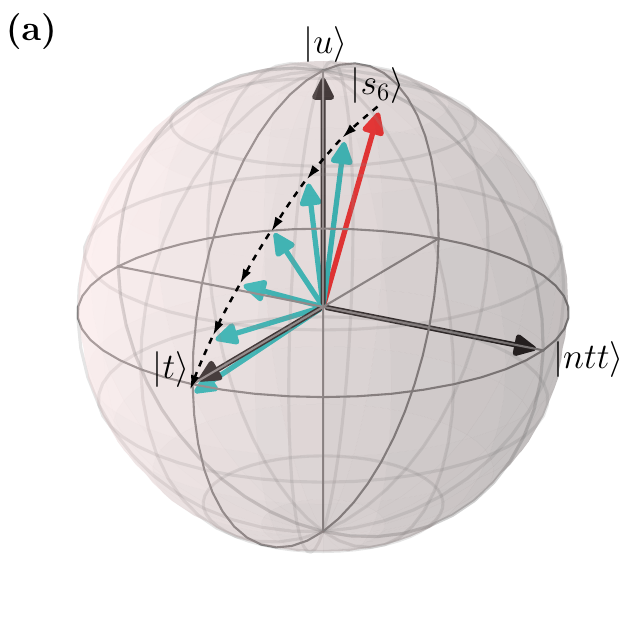}%
}\hfill
\subfloat{%
  \includegraphics[width=0.33\textwidth]{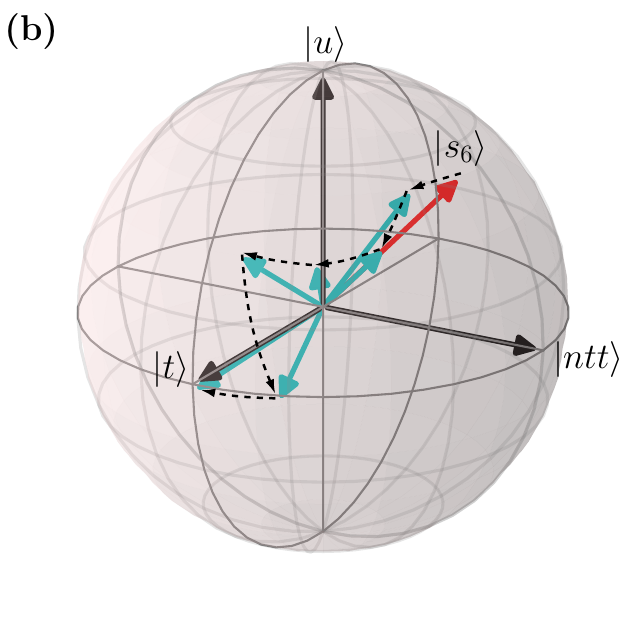}%
}\hfill
\subfloat{%
  \includegraphics[width=0.33\textwidth]{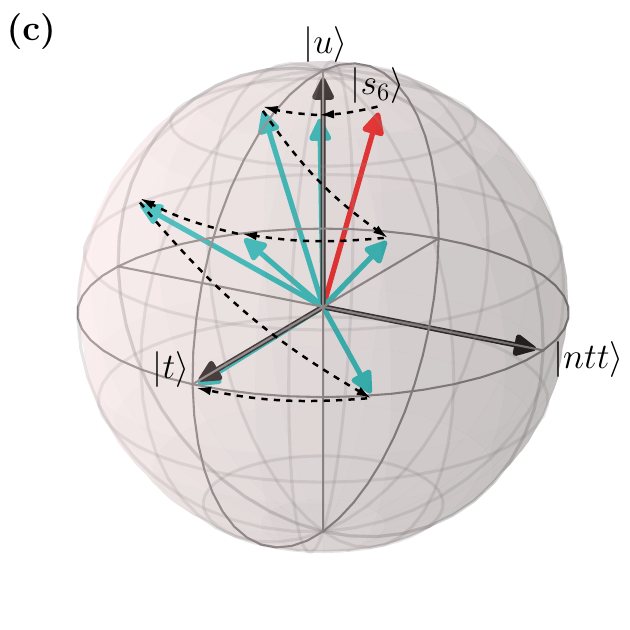}%
}
\caption{Geometric picture of $6$-qubit quantum search algorithm. (a) Grover's algorithm rotates the initial state $|s_6\rangle$ to the target state $|t\rangle$. The total number of oracle operators is $k_\text{opt} = 6$. The final state is not perfectly overlapped with the target state. The success probability is around $99.66\%$. (b) The operator $S_{6,5}(1,1,1,2,1)$ finds the target state with a success probability around $99.86\%$. It has the same number of oracles as Grover's algorithm. (c) The operator $S_{6,3}(1,1,2,1,2)$ finds the target state with a success probability around $99.97\%$. It uses one more oracle operator compared to Grover's algorithm. }\label{fig:6q}
\end{figure*}

The deterministic 4-qubit search algorithm suggests that we can apply the local diffusion operator to design the search algorithm with a higher success probability compared to Grover's algorithm. Suppose that we have a sequence composed of local and global Grover operators. The total number of Grover operators, namely the total number of oracles, is denoted as $k_\text{tot}$. Since $G_n$ and $G_m$ do not commute, there are $2^{k_\text{tot}}$ different operators, and the success probabilities given by them are different. Consider the general form of the sequence
\begin{equation}
\label{eq:S_n_m}
    S_{n,m}(k_1,k_2,\cdots,k_{q-1},k_q) = G_n^{k_1}G_m^{k_2}\cdots G_n^{k_{q-1}}G_m^{k_q}.
\end{equation}
To remove the ambiguity of notation $S_{n,m}$, we set $k_q$ as the number of local Grover operator $G_m$. For example, we have $S_{4,2}(1,1,2) = G_2G_4G_2^2$ and $S_{4,2}(3,0) = G^3_4$. The total number of the oracle operator is 
\begin{equation}
    k_\text{tot} = \sum_{j=1}^q k_j.
\end{equation}
The success probability of finding the target state is
\begin{multline}
    \pr_{n,m}(k_1,k_2,\cdots,k_{q-1},k_q) \\
    = |\langle t|S_{n,m}(k_1,k_2,\cdots,k_{q-1},k_q)|s_n\rangle|^2.
\end{multline}
We restore Grover's algorithm by $S_{n,m}(k_\text{opt},0)$.

We aim to find the operator $S_{n,m}(k_1,k_2,\cdots,k_{q-1},k_q)$ which outperforms Grover's algorithm. There would be an infinite number of sequences if $k_\text{tot}$ is unlimited. To have a fair comparison, we limit to $k_\text{tot} = k_\text{opt}$ or $k_\text{tot} = k_\text{opt}+1$. If $k_\text{tot}<k_\text{opt}$, Grover's algorithm always gives the highest success probability. We exhaustively search for the optimal $S_{n,m}(k_1,k_2,\cdots,k_{q-1},k_q)$ (with the limit $k_\text{tot} = k_\text{opt}$ or $k_\text{tot} = k_\text{opt}+1$) which gives higher success probabilities than Grover's algorithm. We list the results with $n=6,7,8,9$ in Table \ref{tab:one_stage}. Only if Grover's algorithm is overshooting, it is possible to find operators $S_{n,m}(k_1,k_2,\cdots,k_{q-1},k_q)$ better than Grover's algorithm using the same number of oracles. But we can always find operators that improve the success probability with the cost of one more oracle. In particular, we find more than five thousand operators that have higher success probabilities than $9$-qubit Grover's algorithm.

As an illustration, we draw the three-dimensional picture (based on the orthonormal basis $\{|t\rangle,|ntt\rangle,|u\rangle\}$) of 6-qubit Grover's algorithm, and the improved algorithm given by $S_{6,5}(1,1,1,2,1)$ and $S_{6,3}(1,1,2,1,2)$. See Figure \ref{fig:6q}.  Since the local Grover operator $G_m$ is a rotation around the $|u\rangle$ axis, the amplitude of $|u\rangle$ does not change. The global Grover operator $G_n$ is to decrease the amplitude of $|u\rangle$. Note that $\langle u|s_n\rangle = \cos\theta_{n-m}$. Therefore the initial state has a smaller component of $|u\rangle$ if $n$ is larger.

The possible number of sequences $S_{n,m}$ is $2^{k_\text{tot}}$ which scales exponentially with the oracle number $k_\text{tot}$. Therefore, it would be impractical for the classical computer to solve the optimization of $S_{n,m}$ for large $k_\text{tot}$. Note that a large $k_\text{tot}$ means that the size of the database is large (assuming the uniqueness of the target state). Grover's algorithm has a failure probability scale of $1/N$. Therefore, Grover's algorithm for the large database is already near-deterministic. For example, Grover's algorithm has a failure probability $3.1522\times 10^{-6}$ with $n=11$. The room for improvement of the success probability is small. In other words, it is not applicable to optimize $S_{n,m}$ with large $k_\text{tot}$.

\section{\label{sec:two_stage} Two-stage near-deterministic quantum search algorithm}

When the mid-circuit measurements are allowed, we show how to construct the deterministic and near-deterministic search algorithms in Subsections \ref{subsec:two_stage_4q} and \ref{subsec:two_stage} respectively. Similar to the one-stage deterministic or near-deterministic search algorithms, we need classical optimization of the operator sequence. In Subsection \ref{subsec:universal_strategy} we present a strategy improving the success probability without the classical optimization. 

\subsection{\label{subsec:two_stage_4q} Two-stage deterministic \texorpdfstring{$4$}{Lg}-qubit quantum search algorithm}

Because of the nature of the local diffusion operator, we can introduce the mid-measurement and reset in the quantum search algorithm. Recall that we have the deterministic $4$-qubit search algorithm given by Eq. (\ref{eq:4q_deterministic}). The quantum circuit is
$$
\label{circ:4q_deter}
\begin{quantikz}[column sep=0.27cm]
 \lstick{$|0\rangle^{\otimes 2}$} & \qwbundle{2} & \gate{H^{\otimes 2}} & \gate[2]{G_2} & \gate[2]{G_2} & \gate[2]{G_4} & \gate[2]{G_2} & \meter{} \\
 \lstick{$|0\rangle^{\otimes 2}$} & \qwbundle{2} & \gate{H^{\otimes 2}} & & & & & \meter{} 
\end{quantikz}
$$
We set that the local diffusion operator acts on the bottom two qubits. Since the local diffusion operator leaves the top two qubits untouched and the circuit has the success probability $100\%$, it implies that they are already $|t_1\rangle$ (with $|t\rangle=|t_1\rangle\otimes |t_2\rangle$) before the last $G_2$ operator. We can also see that the amplitude of $|u\rangle$ is zero before the last $G_2$ operator. See Fig. \ref{fig:4q}. Therefore, we can measure the top two qubits in advance. 

The bottom two qubits are not in the target state yet if it is not acted by the last $G_2$. However, the $2$-qubit search algorithm is deterministic and requires only one $G_2$ operator. We can disassemble the circuit in Eq. (\ref{circ:4q_deter}) as
$$
\begin{quantikz}[column sep=0.25cm]
 \lstick{$|0\rangle^{\otimes 2}$} & \qwbundle{2} & \gate{H^{\otimes 2}} & \gate[2]{G_2} & \gate[2]{G_2} & \gate[2]{G_4} & \meter{a_1,a_2} \\
 \lstick{$|0\rangle^{\otimes 2}$} & \qwbundle{2} & \gate{H^{\otimes 2}} & & & & 
\end{quantikz}
$$
$$
    \begin{quantikz}[column sep=0.25cm]
        \lstick{$|0\rangle$} & \gate{X^{a_1}} & \gate[3]{G_2} & \\
        \lstick{$|0\rangle$} & \gate{X^{a_2}} & & \\
        \lstick{$|0\rangle^{\otimes 2}$} & \qwbundle{2} & & \meter{} 
    \end{quantikz}
$$
Here $a_1,a_2\in\{0,1\}$ are the measurement results. The first circuit finds $|t_1\rangle$ deterministically. It is a quantum partial search algorithm \cite{GR05}. The second circuit is a normalized $2$-qubit search, which is also deterministic. 

The two-stage algorithm requires additional initialization of the qubits. However, it is more practical for NISQ processors \cite{preskill2018quantum}. The benchmark of the quantum search algorithm on NISQ devices shows that the success probabilities of the quantum search algorithm are not linearly related to the number of two-qubit gates in the circuit \cite{zhang2021implementation}. If the number of two-qubit gates is above a threshold value, the success probability would significantly decrease. We may get no useful information from the one-stage circuit, but the two-stage circuit has a higher chance of providing measurement statistics that distinguish the target states.

\subsection{\label{subsec:two_stage} Two-stage near-deterministic quantum search algorithm with \texorpdfstring{$n\geq 5$}{Lg}}

\begin{table*}[t]
    \caption{Comparison between Grover's algorithm and two-stage near-deterministic search algorithm. The designed operator is limited to $k_\text{tot} = k_\text{opt}+1$ or $k_\text{tot} = k_\text{opt}+2$. The notation $\#$operator means the number of operators giving the success probability higher than Grover's algorithm. Specifically, the numerical criterion is that the success probability is larger than Grover's algorithm at least $10^{-4}\%$.}
    \centering
    \begin{tabular}{c|c|c|c|c|c|c|c|c}
    \hline\hline
         \multirow{2}{*}{$~~n~~$} & \multicolumn{2}{c|}{Grover} & \multicolumn{3}{c|}{Two-stage $k_\text{tot}=k_\text{opt}+1$} & \multicolumn{3}{c}{Two-stage $k_\text{tot}=k_\text{opt}+2$} \\\cline{2-9}
         & $~~k_\text{opt}~~$ & Pr (\%) & Pr (\%) & Operator & ~~\#Operator~~ & Pr (\%) & Operator & ~~\#Operator~~ \\\hline
         5 & 4 & ~~99.91823~~ & ~~99.97864~~ & $S_{5,2}(1,1,1,1)|S_{5,2}(1)$ & 3 & ~~99.98364~~ & $S_{5,2}(2,1,1,1)|S_{5,2}(1)$ & 4 \\
         6 & 6 & 99.65857 & 99.99949 & $S_{6,2}(1,1,3,1)|S_{6,2}(1)$ & 5 & 99.99949 & $S_{6,2}(1,1,1,3,1)|S_{6,2}(1)$ & 12 \\
         7 & 8 & 99.56199 & 99.99993 & $S_{7,2}(1,1,6,0)|S_{7,2}(1)$ & 2 & 99.63717 & $S_{7,2}(1,1,1,1,1,1,3,0)|S_{7,2}(1)$ & 20 \\
         8 & 12 & 99.99470 & 99.99857 & $~~S_{8,2}(1,1,10,0)|S_{8,2}(1)~~$ & 1 & 99.99716 & $S_{8,2}(1,2,1,1,8,0)|S_{8,2}(1)$ & 5 \\
         9 & 17 & 99.94480 & 99.99523 & $S_{9,2}(1,1,15,0)|S_{9,2}(1)$ & 2 & 99.94827 & $~~S_{9,2}(1,1,9,1,1,1,4,0)|S_{9,2}(1)~~$ & 23 \\
    \hline\hline
    \end{tabular}
    \label{tab:two_stage}
\end{table*}

Suppose that the two-stage search algorithm finds the partial solution $t_1$ ($n-m$ bit length) in the first stage. Then the second stage finds the rest of the solution $t_2$ ($m$ bit length). In the framework of the quantum partial search algorithm, the operator $S_{n,m}(k_1,k_2,\cdots,k_{q-1},k_q)$ defined in Eq. (\ref{eq:S_n_m}) finds the partial solution $t_1$ with the success probability
\begin{multline}
    \pr^{(1)}_{n,m}(k_1,k_2,\cdots,k_{q-1},k_q) \\
    = 1-|\langle u|S_{n,m}(k_1,k_2,\cdots,k_{q-1},k_q)|s_n\rangle|^2.
\end{multline}
The superscript $(1)$ means the first-stage algorithm. To take advantage of deterministic $2$-qubit Grover's algorithm, we set $m=2$. Then the second-stage algorithm realized by $G_2$ is deterministic. The overall success probability of finding the full solution is simply $\pr^{(1)}_{n,2}(k_1,k_2,\cdots,k_{q-1},k_q)$. Suppose that the oracle number of the first-stage algorithm is $k'_\text{tot}$. Then the overall oracle number is $k_\text{tot}=k'_\text{tot}+1$. We can also set $m=4$, and design the second-stage by $S_{4,2}(1,1,2)$. The corresponding efficiency is $k_\text{tot}=k'_\text{tot}+4$.

To compare with Grover's algorithm, the total number of oracle operators is set to $k_\text{tot}=k_\text{opt}+1$ or $k_\text{tot}=k_\text{opt}+2$. We exhaustively search for the optimal operators for $5\leq n\leq 9$. There always exist operators which give higher success probabilities than Grover's algorithm. See Figure \ref{fig:success_probability} and Table \ref{tab:two_stage}. The success probabilities are all above $99.9\%$, therefore near-deterministic. Because of the constrain $k_\text{tot}=k_\text{opt}+1$ or $k_\text{tot}=k_\text{opt}+2$, we find that the second-stage algorithm is always the deterministic $2$-qubit search rather than the deterministic $4$-qubit search. Note that the operators $S_{7,2}(1,1,6,0)$, $S_{8,2}(1,1,10,0)$, and $S_{9,2}(1,1,15,0)$ represent the standard quantum partial search algorithm, namely the global-local-global operator order \cite{GR05,KG06,Korepin05}. While other cases are non-standard operators but can find the partial solution with success probabilities close to 1.

\subsection{\label{subsec:universal_strategy} Universal strategy of improving the success probability}

The second stage of the algorithm is fixed as the $2$- or $4$-qubit deterministic search algorithm. To achieve the maximal success probability, the first-stage algorithm is designed through classical optimization. However, we can always set the first-stage operator as $S_{n,m}(k_\text{tot},0)$ with $m=2$ or $m=4$, namely the standard Grover operator. The difference is that we use the standard Grover operator for the partial search (finding $n-m$ bits of the solution). Notice that the normalized all non-target state $|t^\perp\rangle$ defined in Eq. (\ref{eq:t_perp}) can be rewritten as
\begin{equation}
    |t^\perp\rangle = \frac{\sin\theta_{n-m}\cos\theta_m}{\cos\theta_n}|ntt\rangle + \frac{\cos\theta_{n-m}}{\cos\theta_n}|u\rangle.
\end{equation}
Based on Eq. (\ref{eq:G_n^k}), the $k_\text{opt}$ iteration of $G_n$ gives
\begin{multline}
    G_n^{k_\text{opt}}|s_n\rangle = \sin\left((2k_\text{opt}+1)\theta_n\right)|t\rangle \\
    + \cos\left((2k_\text{opt}+1)\theta_n\right)\frac{\sin\theta_{n-m}\cos\theta_m}{\cos\theta_n}|ntt\rangle \\
    + \cos\left((2k_\text{opt}+1)\theta_n\right)\frac{\cos\theta_{n-m}}{\cos\theta_n}|u\rangle.
\end{multline}
Then the failure probability of finding $n-m$ bits of the solution is 
\begin{equation}
    1- \pr^{(1)}_{n,m}(k_\text{opt},0) = \cos^2\left((2k_\text{opt}+1)\theta_n\right)\frac{\cos^2\theta_{n-m}}{\cos^2\theta_n}.
\end{equation}
Since the second stage is deterministic if $m=2$ or $m=4$, then the above is also the failure probability of the full search. Grover's algorithm has the failure probability
\begin{equation}
     1- \pr_{n,m}(k_\text{opt},0) = \cos^2\left((2k_\text{opt}+1)\theta_n\right).
\end{equation}
Compare the above probabilities, then we have the improvement 
\begin{equation}
    \Delta\pr = 1- \frac{\cos^2\theta_{n-m}}{\cos^2\theta_n} = \frac{2^m-1}{2^n-1}.
\end{equation}
If $m=2$, we improve the success probability by $3/(2^n-1)$ with the cost of one more $G_2$. If $m=4$, we improve the success probability by $15/(2^n-1)$ with the cost of four more Grover operators (rescaled deterministic $4$-qubit search algorithm).

The improved success probability comes from using the global Grover operator as a partial search. The increased success probability scales as $\mathcal O(2^{-(n-m)})$. It has no quadratic speed-up. It suggests that the partial diffusion operator works better for the partial search algorithm. In fact, the increased success probability can scale as $\mathcal O(2^{-(n-m)/2})$ if the partial diffusion operator is allowed. From Table \ref{tab:two_stage}, we can see that the optimal first-stage $S_{n,m}$ always includes some partial diffusion operators.

\section{\label{sec:multi-target}Discussions on multi-target search problem}

We assume the uniqueness of the solution in the database. Grover's algorithm can find one of the target states if there are $M$ solutions. Suppose that the ratio is $\lambda = M/N$. The optimal iteration number becomes $k_\text{opt} = \lfloor \pi/(4\theta_\lambda)-1/2\rceil$ with $\theta_\lambda = \arccos(1/\sqrt{\lambda})$. The deterministic search algorithm via the delicately designed phase can also be easily generalized to the multi-target cases \cite{long2001grover,roy2022deterministic,leng2023improving}. The phases are also related to the ratio $\lambda$. 

If there are multiple target states, the action of the local diffusion operator becomes complicated. Even with the same operator, different choices of the qubits (acted by the local diffusion operator) give different amplitudes. If the $M$ target states are evenly distributed in the target blocks (each target block has the same number of target states), we have a simple generalization of the quantum partial search algorithm \cite{choi2007quantum}. The three-dimensional geometric picture is still valid. Therefore, our designed near-deterministic algorithm works with simple modifications. 

If the $M$ target states are unevenly distributed in the target blocks, the global Grover operator $G_n$ becomes elements of $O(2l+1)$ group with $l$ target blocks \cite{zhang2018quantum}. The number of target states in each target block is needed to design the proper partial search. Similar rules are applied to design the multi-target near-deterministic search algorithm. The optimal operator which gives the maximal success probability is dependent on the distribution information of the target states. There may exist multi-target deterministic one-stage or two-stage search algorithms without phase design. We leave it for future study.

\section{\label{sec:conclusion} Conclusion}

We design the near-deterministic search algorithm with the help of the local diffusion operator. It does not require delicate design of the phases in the oracle or the diffusion operator. We show that the success probability of Grover's algorithm can be improved with an additional one or two more Grover operators. 

The implementations of the local diffusion operator require fewer circuit depths compared to the global one \cite{zhang2020depth}. Therefore our near-deterministic search algorithm may be more efficient compared to Grover's algorithm (even with additional Grover operators). Another advantage is to allow the divide-and-conquer strategy (two-stage search algorithm). Therefore the algorithms may be more friendly to NISQ devices. The local diffusion operator may be also useful in designing the fixed-point search algorithm \cite{yoder2014fixed}. Whether our algorithm can be generalized to the amplitude amplification algorithm \cite{grover1998quantum,brassard2002quantum} is another interesting question worth exploring in the future.

\begin{acknowledgments}

The work of K.Z. was supported by the National Natural Science Foundation of China under Grant Nos. 12305028 and 12247103, and the Youth Innovation Team of Shaanxi Universities.

\end{acknowledgments}


\providecommand{\noopsort}[1]{}\providecommand{\singleletter}[1]{#1}%

\end{document}